\documentstyle[psfig,preprint,aps,eqsecnum,floats]{revtex}

\newcommand{\beq}{\begin{equation}}
\newcommand{\eeq}{\end{equation}}
\newcommand{\beqa}{\begin{eqnarray}}
\newcommand{\eeqa}{\end{eqnarray}}

\def\epsK{\varepsilon_K}
\def\O{{\cal O}}
\def\aPK{a_{\psi K_S}}
\def\aPP{a_{\pi\pi}}
\def\dmd{\Delta m_{B_d}}
\def\dms{\Delta m_{B_s}}

\begin{document}

{\tighten
\preprint{\vbox{\hbox{SLAC-PUB-7622}
                \hbox{WIS-97/27/Sep-PH}
                \hbox{LAL 97-65}
                \hbox{hep-ph/9709288}}}

\title{Implications of the Fleischer-Mannel Bound}
\author{Yuval Grossman$^a$, Yosef Nir$^b$,
St\'ephane Plaszczynski$^c$ and Marie-H\'el\`ene Schune$^c$}

\address{ \vbox{\vskip 0.truecm}
$^a$Stanford Linear Accelerator Center \\
        Stanford University, Stanford, CA 94309, USA \\
        {\rm yuval@slac.stanford.edu} \\
\vbox{\vskip 0.truecm}
  $^b$Department of Particle Physics \\
  Weizmann Institute of Science, Rehovot 76100, Israel \\
        {\rm ftnir@weizmann.weizmann.ac.il} \\
\vbox{\vskip 0.truecm}
  $^c$Laboratoire de l'Acc\'el\'erateur Lin\'eaire \\
 IN2P3-CNRS et Universit\'e de Paris-Sud, F-91405 Orsay, France \\
         {\rm plaszczy,schunem@lal.in2p3.fr}
 }
\maketitle
\begin{abstract}%
\noindent
Fleischer and Mannel (FM) have shown that it may become possible to
constrain the angle $\gamma$ of the unitarity triangle from measurements
of various $B\to\pi K$ decays. This constraint is independent of hadronic
uncertainties to the few percent level. We show that, within the
Standard Model, the FM bound gives strong constraints on the CKM
parameters. In particular, it could predict a well defined sign
for $\sin2\gamma$ and $\sin2\alpha$. In a class of extensions
of the Standard Model, where the New Physics affects only
$\Delta B=2$ (and, in particular, not $\Delta B=1$)
processes, the FM bound can lead to constraints
on CP asymmetries in $B$ decays into final CP eigenstates
even if $B-\bar B$ mixing is dominated by unknown New Physics.
In our analysis, we use a new method to combine in a statistically
meaningful way the various measurements that involve CKM parameters.

\end{abstract}
}
\newpage

\section{Introduction}

Fleischer and Mannel \cite{fm} have shown that, using the branching
ratios of four $B\to\pi K$ decay modes, it is possible to derive a bound
on the angle $\gamma$ of the unitarity triangle which,
under certain circumstances, is free of hadronic uncertainties.
In this work we show that this bound can provide strong constraints
on the CKM parameters within the standard model as well as
model independent predictions for various CP asymmetries in neutral
$B$ decays.

CKM unitarity allows one to describe any $B$ decay amplitudes as a sum
of two terms, each with a definite weak phase related to a particular
combination of CKM-matrix elements \cite{gq}.  For $b \to q \bar q s$
decays, it is convenient to choose the two terms as
$A=A_c+A_ue^{-i\gamma}e^{i\delta}$, where $A_c \propto |V_{cb}V_{cs}|$,
$A_u \propto |V_{ub}V_{us}|$, $\gamma$ is the CP violating angle of the
unitarity triangle \cite{rev} and $\delta$ is a CP conserving
strong phase.  The amplitudes for the relevant $B\to\pi K$ decays are
then written as follows:
\beqa \label{fourmodes}
A(B^0 \to \pi^- K^+) = A_c^0 - A_u^0 e^{i\gamma} e^{i\delta}, &\qquad&
A(\bar B^0 \to \pi^+ K^-) = A_c^0 - A_u^0 e^{-i\gamma} e^{i\delta},
 \nonumber \\
A(B^+ \to \pi^+ K^0) = A_c^+ - A_u^+ e^{i\gamma} e^{i\delta'},&\qquad&
A(B^- \to \pi^- \bar K^0) = A_c^+ - A_u^+ e^{-i\gamma} e^{i\delta'},
\eeqa
The following two assumptions are very likely to hold with regard
to these four channels:

{\it 1. The contributions to $A_u$ that do not come from tree-level
amplitudes can be neglected} \cite{Flei}.
The reason is that the
penguin amplitudes contributions to $A_u$ are suppressed compared
to their contributions to $A_c$ by $\O(|V_{ub}V_{us}|/|V_{tb}V_{ts}|)
\sim0.02$. Then in the charged $B$ decays, which require a
$b\to d\bar ds$ transition, we can neglect $A_u$ while in the neutral $B$
decays, which are mediated by a $b\to u\bar u s$ transition, we
take into account only the tree-level amplitude $A_T$:
\beq \label{AuAT}
A_u^+=0,\ \ \ A_u^0=A_T.
\eeq

{\it 2. The contributions from electroweak penguins can be neglected}
\cite{Flei}. Indeed these contributions can be reliably estimated and
they are expected to be of $\O(0.01)$ of the leading contributions.
Then $A_c$ comes purely from QCD penguin amplitudes $A_P$
which, as a result of the $SU(2)$ isospin symmetry of the strong
interactions, contribute equally to the charged and neutral $B$ decays:
\beq \label{AcAP}
A_c^0 = A_c^+ = A_P.
\eeq

With the two approximations
(\ref{AuAT}) and (\ref{AcAP}) one gets \cite{fm}
\beqa
\Gamma (B_d \to \pi^\mp K^\pm) &\equiv &
{\Gamma (B^0 \to \pi^- K^+) + \Gamma (\bar B^0 \to \pi^+ K^-) \over 2}
\propto |A_P|^2 (1-2r\cos\gamma\cos\delta + r^2), \nonumber \\
\Gamma (B^\pm \to \pi^\pm K) &\equiv &
{\Gamma (B^+ \to \pi^+ K^0) + \Gamma (B^- \to \pi^- \bar K^0) \over 2}
\propto |A_P|^2,
\eeqa
where
\beq
r\equiv A_T/A_P.
\eeq
This leads to
\beq \label{aa}
R \equiv
{\Gamma (B_d \to \pi^\mp K^\pm) \over \Gamma (B^\pm \to \pi^\pm K)} =
1-2r\cos\gamma\cos\delta + r^2.
\eeq
It is clear that $R$ can be smaller than 1 only if there is
a destructive interference between the penguin and tree contributions
in the neutral $B$ decays. This requires that both $\cos\gamma$ and
$\cos\delta$ do not vanish. Thus, if $R<1$ we may get some useful
information on $\gamma$.

In general, the constraints on $\gamma$ will depend on hadronic
physics. In particular, while $R$ is a measurable quantity, $r$
and $\cos\delta$ are hadronic, presently unknown parameters.
(We treat $r$ as a free parameter. Estimates based on factorization
and on $SU(3)$ relations prefer $r\lesssim 0.5$ \cite{fm}.)
Fortunately, for a given value of $\cos\gamma\cos\delta$, $R$ has
a minimum value as a function of $r$. To find this minimum, we solve
\beq
{dR \over dr} = -2 \cos \gamma \cos\delta + 2r = 0,
\eeq
which leads to $R_{\rm min}=R(r=\cos\gamma\cos\delta)$, namely
\beq
R \ge 1- \cos^2 \gamma \cos^2\delta.
\eeq
The Fleischer-Mannel (FM) bound is derived by setting $\cos\delta=1$:
\beq\label{TheLimit}
\sin^2\gamma \le R.
\eeq
Note that a similar bound for $\delta$, $\sin^2\delta\le R$, can be
obtained. Also note that additional decay modes, such as $B\to\pi K^*$
and $B\to\rho K$, can be used for this analysis.

Clearly, the bound (\ref{TheLimit}) is significant only for $R<1$,
as explained in \cite{fm}. Recent CLEO results \cite{CLEO}\ give
\beq \label{CleoRange}
R=0.65\pm0.40.
\eeq
Thus, we may be fortunate and indeed have $R<1$. As soon as
an upper bound on $R$ below unity is obtained, the limit
(\ref{TheLimit}) will give useful constraints in the $\rho-\eta$ plane
within the Standard Model and in the $\aPP-\aPK$ (the CP asymmetries in
$B\to\pi\pi$ and $B\to\psi K_S$, respectively%
\footnote{By $\aPP$ we refer to the CP asymmetry in the
$W$-mediated tree-level decay. Isospin analysis will, very likely, be
needed to eliminate the `penguin pollution' \cite{GrLoIs}. $\aPP$ can
also be deduced from the CP asymmetry in $B\to\rho\pi$ combined with
isospin analysis \cite{LNQS,Gron,QuSn}.})
plane for a class of extensions of the Standard Model
\cite{GNW}. We now describe the derivation and significance of these
constraints.

\section{Standard Model Analysis}

Within the Standard Model, bounds on the CKM parameters are
often presented as constraints on the unitarity triangle in the
$\rho-\eta$ plane. In Fig. 1, we show the present bounds from
$|V_{cb}|$, $|V_{ub}/V_{cb}|$, $\Delta m_{B_d}$, $\epsK$ and
$\Delta m_{B_s}$ (see the Appendix for a detailed explanation
of our method in combining the constraints). The limit (\ref{TheLimit})
translates into an exclusion region in this plane:
\beq
{\eta^2\over\rho^2+\eta^2}\le R.
\eeq

Examples of the exclusion regions are shown in Fig. 2. Once the upper
bound on $R$ is below 1, a region around $\rho=0$ is excluded.
The choice of these examples is based on the following naive
scaling arguments. The CLEO result (\ref{CleoRange}) was obtained with
about 3.3 fb$^{-1}$. By the beginning of the $B$-factories era, CLEO
should reach about 10 fb$^{-1}$, so a gain of $\sqrt{3}$ on the
statistical error is expected. This gives $\sigma_R=0.22$ which,
for a central value of $R=0.65$, has still only a small effect compared
to the allowed region of Fig. 1. After one year of CLEOIII, BaBar and
BELLE we could have about 80 fb$^{-1}$, so a gain
of about a factor of 5 on the error is expected, namely $\sigma_R=0.08$.

Another useful presentation is in the $\sin2\alpha-\sin2\beta$ plane
\cite{SoWo,NiSa}. The present allowed region at 95\% CL is shown in Fig.
3. Since $\sin^2\gamma=1$ corresponds to $\sin2\alpha=\sin2\beta$, once
the upper bound on $R$ is below 1, a region around the $\sin2\alpha
=\sin2\beta$ line is excluded. Examples of such constraints are depicted
in Fig. 4.

We would like to point out two potentially interesting situations
which might develop in the future.

First, the combination of a lower bound on $B_s-\bar B_s$ mixing
and an upper bound on $R$ may be very powerful in excluding
the possibility of a negative $\cos\gamma$. The reason is that
the $\dms/\dmd$ bound puts a lower bound on $\cos\gamma$ while
the $R$ bound translates into an upper bound on negative $\cos\gamma$.
To see this explicitly, let us define an $SU(3)$ breaking factor
\beq
R_{SU(3)}=\left({m_{B_s}\over m_{B_d}}\right)
\left({B_{B_s}f_{B_s}^2\over B_{B_d}f_{B_d}^2}\right).
\eeq
Then, $|V_{td}/V_{ts}|^2=R_{SU(3)}(\dmd/\dms)$ gives
\beq \label{sdgamma}
\cos\gamma\ge{|V_{us}|^2+|V_{ub}/V_{cb}|^2-R_{SU(3)}(\dmd/\dms)_{\rm max}
\over 2|V_{us}||V_{ub}/V_{cb}|} \ge 1.97-{27.5\;{\rm ps}^{-1}\over\dms},
\eeq
where we use $|V_{us}|=0.22$, $\dmd<0.481\;$ps$^{-1}$ \cite{Schn},
$|V_{ub}/V_{cb}|>0.06$ \cite{PDG}\ and $R_{SU(3)}<1.51$ \cite{Sach} to
get the second inequality. On the other hand, the bound (\ref{TheLimit})
gives an upper bound on $\cos\gamma$ if $\cos\gamma$ is negative:
\beq \label{negcos}
\cos\gamma\le-\sqrt{1-R}\ \ {\rm for}\ \cos\gamma<0.
\eeq

Eq. (\ref{sdgamma}) implies that if $(\dms)_{\rm min}\ge14.0\;$ps$^{-1}$
is reached, then the $\dms$ bound by itself will be
enough to exclude negative $\cos\gamma$. But for the interesting range
between the present 95\% CL lower bound \cite{Schn}, $(\dms)_{\rm min}=
10.0\;$ps$^{-1}$, and $(\dms)_{\rm min}=14.0\;$ps$^{-1}$, only the
combination with a low enough $R_{\rm max}$ can exclude the negative
$\cos\gamma$ range. For example, with $(\dms)_{\rm min}=10.0\;$ps$^{-1}$,
eq. (\ref{sdgamma}) gives $\cos\gamma\ge-0.78$, while
$R\le0.39$ allows negative $\cos\gamma$ only below --0.78.
The combination of $\dms\ge10.0\;$ps$^{-1}$ and $R\le0.39$ excludes then a
negative $\cos\gamma$ and actually allows only $\cos\gamma\ge0.78$.
The $R_{\rm max}$ values required to close the
negative $\cos\gamma$ window for various $(\dms)_{\rm min}$ values are
given in Table 1.\footnote{In our calculations, as explained in the
Appendix, we use the full experimental information on $\dms$ and not just
the lower bound, so that a negative $\cos\gamma$ can be excluded by
somewhat weaker bounds on $R$.}
Of course, a stronger lower bound on $|V_{ub}/V_{cb}|$ (above 0.06)
and/or a stronger theoretical upper bound on $R_{SU(3)}$ (below 1.51)
will make the task of closing the negative $\cos\gamma$ window easier.

\begin{table}
\[ \begin{array}{||c||c|c|c|c||} \hline\hline
(\dms)_{\rm min}\ \left[{\rm ps}^{-1}\right] & 10 & 11 & 12 & 13 \\
\hline
R_{\rm max}           & 0.39 & 0.72 & 0.90 & 0.98 \\ \hline
(\cos\gamma)_{\rm min} & 0.78 & 0.53 & 0.32 & 0.15 \\ \hline\hline
\end{array} \]
\label{Rdms}
\vspace{0.1in}
\caption
{Lower bounds on $\dms$ and upper bounds on $R$ that close the negative
$\cos\gamma$ window and the corresponding lower bound on $\cos\gamma$.}
\end{table}

Second, the above $\dms-R$ combination, together with the existing strong
constraints on $\beta$, can exclude a positive $\sin2\alpha$.
The bound $|V_{ub}/V_{cb}|\le0.10$ gives $\beta<0.15\pi$.
Suppose that $R<0.79$ is established and, furthermore,
$\dms$ is known to be large enough that the negative $\cos\gamma$
window is closed (this would happen under these circumstances
with $\dms>11.3\;$ps$^{-1}$).
Then we will get a lower bound $\cos\gamma>0.46$ which
is equivalent to $\gamma<0.35\pi$. Together with the upper bound on
$\beta$, we get $\alpha>\pi/2$, namely $\sin2\alpha<0$.

To summarize: a combination of (i) a range for $|V_{ub}/V_{cb}|$,
(ii) a lower bound on $\dms$, (iii) an upper bound on $R$ and (iv)
the information from $\epsK$ that $\eta>0$, might exclude large regions
in the $\rho-\eta$ and $\sin2\alpha-\sin2\beta$ planes that are presently
allowed. An example of the above situations is given in Figs. 2(b) and
4(b) where an improved measurement for $R$ is assumed. We have a clear
prediction of $\cos\gamma>0$ (see Fig. 2(b)) and $\sin2\alpha<0$
(see Fig. 4(b)).

\section{Beyond the Standard Model}

We now turn to a discussion of the implications of the FM bound
for theories beyond the Standard Model. If new physics affects
the $B\to\pi K$ decay rates of eq. (\ref{fourmodes}), then the
resulting bound (\ref{TheLimit}) might be in conflict with
other CKM constraints, thus probing this new physics \cite{fmtwo}.
In this work, we focus on extensions of the Standard Model where
the four decay modes (\ref{fourmodes}) are dominated by the
Standard Model diagrams. Yet, we allow for large, even dominant,
contributions from New Physics to $B-\bar B$ mixing and to $\epsK$.
This class of models (without any assumptions on New Physics in
$B\to\pi K$ decays) was studied in ref. \cite{GNW}. It was shown
there that combining the information from the CP asymmetries in
$B\to\psi K_S$ ($\aPK$) and in $B\to\pi\pi$ ($\aPP$) with the measurement
of $|V_{ub}/V_{cb}|$ allows one to reconstruct the unitarity
triangle.\footnote{%
For $\aPP$ one can combine measurements of various decays that are
dominated by the $b\to u\bar ud$ transition such as the $2\pi$, $3\pi$
and $4\pi$ final states. For $\aPK$ one can combine measurements of
various decays that are dominated by the $b\to c\bar cd$ and
$b\to c\bar cs$ transitions such as
the $\psi K_S$, $\psi K^*$ and $D^+D^-$ final states.}
Obviously, the FM bound can test this construction. But it also
gives a completely new aspect in the model independent analysis
by predicting correlations between $\aPP$ and $\aPK$.
In particular, it might forbid regions in the
$a_{\psi K_S}-a_{\pi\pi}$ plane. No such definite constraint
arises from the $|V_{ub}/V_{cb}|$ bound alone, which is the
only other CKM constraint that is viable in a large class of
models of new physics.

Let us first repeat the basis for the model independent analysis
\cite{GNW}. We study extensions of the Standard Model with
arbitrary (within, of course, phenomenological constraints)
new physics contributions to $B-\bar B$ mixing and to $K-\bar K$ mixing.
On the other hand, we assume that the following features hold:
\begin{description}
\item[$(i)$]
{The $\bar b\to\bar cc\bar s$ and $\bar b\to\bar uu\bar d$
decays for $\aPK$ and $\aPP$ respectively, as well as the semileptonic
$B$ decays for the $|V_{ub}/V_{cb}|$ measurement, are dominated by
Standard Model tree level diagrams.}
\item[$(ii)$]
{Unitarity of the three generation CKM matrix is practically maintained.}
\end{description}
Then, it is possible to use the measurements of $\aPK$, $\aPP$
and $|V_{ub}/V_{cb}|$ to construct the Unitarity Triangle and,
in particular, to determine the angle $\gamma$ up to an eightfold
discrete ambiguity \cite{GNW}. The validity of these ingredients
in extensions of the Standard Model was discussed in \cite{GNW}.
An example of model independent constraints in the $\rho-\eta$ plane
is shown in Fig. 5(a). The derivation of the allowed regions
is explained in the Appendix.

The FM bound provides a constraint on $\gamma$ and therefore
is very interesting for a model independent analysis. However,
to apply it in this analysis, one has to make one further assumption:
\begin{description}
\item[$(iii)$]
{The $\bar b\to\bar uu\bar s$ and $\bar b\to\bar dd\bar s$
decays for the $B\to\pi K$ decays of (\ref{fourmodes})
are dominated by Standard Model diagrams.}
\end{description}
We emphasize that this assumption holds much less generically
than assumption ($i$) above. While ($i$) concerns decays that are
dominated by Standard Model tree diagrams, ($iii$) concerns decays that
are dominated by $b\to s$ penguin transitions. The latter are suppressed
by loop factors and small CKM factors in the Standard Model and thus are
more sensitive to New Physics. Calling the analysis
below `model-independent' might be somewhat misleading in this sense.
Yet, there is a reasonably large class of models where our three
assumptions hold while simultaneously allowing for interesting effects
in $B-\bar B$ mixing. For example, in models with extra down quark
singlets, there could be large CP violating contributions to $B-\bar B$
mixing \cite{NiSi}, while the contributions to $b\to s$ transitions are
constrained by the $B\to\mu^+\mu^- X$ and $B\to\nu \bar\nu X$
bounds to be small \cite{Silv,Bran,gln}.\footnote{CKM unitarity is
violated in this class of models but the effect is small
\cite{NiSi}.}

To make things clear we state again that the following analysis
applies only to models where the three assumptions $(i)-(iii)$ hold.
This is only a subclass of the models to which the analysis of
\cite{GNW}\ applies.

Examining Fig. 5(a), we learn that the FM bound can test the assumptions
that underlie the model independent analysis. A very strong upper bound
on $\sin^2\gamma$ may turn out to be inconsistent with any of the
eight solutions for $\gamma$, implying that there is new physics
in at least some of the relevant $\Delta B=1$ processes. In other
cases the FM bound can be useful in reducing the discrete ambiguity
to fourfold. An example of such a situation is given in Fig. 5(b).

The line of thought that stands in the basis of \cite{GNW}\ can
be taken a step further: if the angle $\gamma$ of the unitarity triangle
is known or, at least, constrained by experimental data, then the
predictions for the CP asymmetries $\aPK$ and $\aPP$ will be correlated.

The $|V_{ub}/V_{cb}|$ measurement does not constrain $\gamma$.
Therefore, the analysis of \cite{GNW}\ could not predict any
correlations between $\aPK$ and $\aPP$: the whole plane (between
$-1$ and $+1$ for each asymmetry) is allowed.
But the FM bound does constrain $\gamma$.
Given an angle $\gamma$ of the unitarity triangle, the crucial
relation in the model independent analysis of \cite{GNW}\ is
\beq
2\gamma+\arcsin(\aPK)+\arcsin(\aPP)=2\pi({\rm mod}\ 4\pi),
\eeq
which is translated in a straightforward way to the following
relation between $\sin^2\gamma$ and the two asymmetries:
\beq \label{gammaab}
(\aPK+\aPP)^2+\tan^2\gamma(\aPK-\aPP)^2=4\sin^2\gamma.
\eeq
Eq. (\ref{gammaab}) defines an ellipse in the $\aPP-\aPK$ plane.
The principal axes are on the diagonals, and the ratio between them is
$|\tan\gamma|$.

An upper bound on $\sin^2\gamma$, such as (\ref{TheLimit}), excludes then
a region in the $\aPP-\aPK$ plane. The excluded region is the area
between the ellipse and the boundaries of the plane close to the
$(+1,+1)$ and  $(-1,-1)$ corners. To understand this picture, one can
think in the following way: for $\sin^2 \gamma=0$, eq. (\ref{gammaab})
gives the diagonal from  $(-1,+1)$ to  $(+1,-1)$. As $\sin^2\gamma$
increases, the diagonal turns into an ellipse with the ratio between the
principal axes, $\tan\gamma$, increasing from $0$ to $\infty$. This
corresponds to the ellipse deforming within the plane.
At $\sin^2\gamma=1$, eq. (\ref{gammaab}) gives the diagonal from
$(-1,-1)$ to  $(+1,+1)$. If we have a bound $\sin^2\gamma<1$, the ellipse
in its deformation does not cover the upper-right and lower-left corners.
An example of the exclusion regions is shown in Fig. 6.

A very interesting constraint on the allowed CP asymmetries
arises in models where a fourth assumption holds:
\begin{description}
\item[$(iv)$]
{CP violation in the neutral kaon system is dominated by
the Standard Model box diagrams. In other words, $\epsK$ is
accounted for by the CKM phase.}
\end{description}
This is an interesting situation because, in this case, $\epsK$
gives a lower bound on $\sin^2\gamma$. This excludes yet another
region in the $\aPP-\aPK$ plane. The excluded region is the area
between the ellipse (\ref{gammaab}) that corresponds to
$(\sin^2\gamma)_{\rm min}$ and the boundaries of the plane close to the
$(-1,+1)$ and  $(+1,-1)$ corners. This should be intuitively clear
from our discussion above of the FM bound in this context.
Taking $B_K<1$, $|V_{cb}|<0.043$ and $|V_{ub}/V_{cb}|<0.10$,
the bound from $\epsK$ reads $\sin\gamma>0.3$.
(The information that is relevant to correlating the asymmetries
through (\ref{gammaab}) is $\sin^2\gamma>0.1$. The fact that $\epsK$
excludes negative $\sin\gamma$ is irrelevant here.)

The combination of upper and lower bounds on $\sin^2\gamma$
(for example the FM bound and the $\epsK$ bound) is even more powerful.
If neither $\sin^2\gamma=1$ nor $\sin^2\gamma=0$ are allowed, then
the ellipse in its deformation 
does not reach not only the corners but
also the origin $(0,0)$. Consequently, in addition to the
areas excluded separately by each of the bounds, also the area around
$(0,0)$ that is inside the overlap of the respective ellipses is
excluded. An example of the three regions is given in Fig. 6.

In ref. \cite{GNW}, two more possible scenarios were examined:
\begin{description}
\item[$(iv)^\prime$]
{The $K_L\to\pi\nu\bar\nu$ decay is dominated by
the Standard Model diagrams.}
\item[$(iv)^{\prime\prime}$]
{The ratio $\Delta m_{B_s}/\Delta m_{B_d}$
corresponds to $R_{SU(3)}|V_{ts}/V_{td}|^2$, even though each
of the two mixing parameters is affected by new physics.}
\end{description}
Each of $(iv$), ($iv)^\prime$ and ($iv)^{\prime\prime}$
holds in some class of models. For certain models, more than one
of these assumptions might hold. In any case, the important
feature for our analysis is that under any of the three
assumptions, future measurements might constrain $\sin^2\gamma$.
Particularly useful will be a lower bound on $\sin^2\gamma$ which
can be combined with the FM bound as explained above. Such a lower
bound exists already for $\epsK$ and can be achieved with a lower
bound on $BR(K_L\to\pi\nu\bar\nu)$ or if $\Delta m_{B_s}$ is measured.
An upper bound on $\sin^2\gamma$ from any of these three measurements
(or bounds)
is also interesting as it will allow an analysis similar to that
of the FM bound within the corresponding class of models.

Finally we note that if any other method to measure or constrain
$\gamma$ becomes available, and if the relevant processes are
dominated by Standard Model contributions in a class of new physics
models, then the above analysis could be applied in a similar way.
In particular, Atwood, Dunietz and Soni \cite{ADS} have recently
proposed an improvement of the Gronau-Wyler \cite{GrLo,GrWy,Duni} method
to measure $\gamma$ via $B\to KD^0(\bar D^0)$ decays. The new method
is not only theoretically clean but might also be experimentally
feasible. The relevant quark transitions are $b\to c\bar us$ and
$b\to \bar cus$ which are dominated by the Standard Model tree diagrams
in many new physics models (see examples in \cite{GrWo}).
Similarly, various proposed methods to measure $\gamma$ through
$B_s$ decays (see {\it e.g.} \cite{FlDu}) can be subject to a similar
analysis.

\section{Summary}

Fleischer and Mannel have suggested a method which,
under certain circumstances, can give an upper bound on $\sin^2\gamma$
that is almost free of hadronic uncertainties. We have shown that within
the Standard Model such a bound can give strong constraints on the
unitarity triangle. While at present the information
from $\epsK$, $\dmd$ and $|V_{ub}/V_{cb}|$ constrains only $\beta$
to be within one quadrant, the addition of the FM and $\dms$ bounds can
potentially constrain each of $\gamma$ and $\alpha$ to a single quadrant.
In particular, if $\sin2\alpha<0$, this can be deduced from improved FM
and $\dms$ bounds (while currently $\sin 2\alpha$ can assume any value).

In extensions of the Standard Model where the New Physics contributes
significantly to $B-\bar B$ mixing but to none of the relevant decay
processes, the FM bound can give correlations between the
allowed values of the CP asymmetries in $B\to\psi K_S$ and
$B\to\pi\pi$. If also a lower bound on $\sin^2\gamma$
is available, for example in models where
$\epsK$ is dominated by the Standard Model,
the excluded region for these asymmetries is very significant.

\acknowledgments
We thank A. Buras, M. Danilov, T. Mannel and M. Worah for discussions
and comments on the manuscript.
Y.G. is supported by the Department of Energy under contract
DE-AC03-76SF00515. Y.N. is supported in part by the United States --
Israel Binational Science Foundation (BSF), by the Israel Science
Foundation, and by the Minerva Foundation (Munich).

\appendix
\section{Fitting the CKM Parameters}
\subsection{Description of the Basic Method}
We explain here our method of statistically combining many measurements
involving CKM parameters \cite{BBN}. The method described below
was adopted by the BaBar collaboration \cite{BaB}.
In this work, we combine existing measurements of $|V_{cb}|$,
$|V_{ub}/V_{cb}|$, $\Delta m_{B_d}$, $\epsK$, and (the lower
bound on) $\Delta m_{B_s}$ with future measurements of the ratio
$R$ defined in eq. (\ref{aa}).

There are two types of errors which enter the determination of the CKM
parameters: experimental errors and uncertainties due to theoretical
model dependence. These two types of errors will be treated differently.

Experimental errors are generally assumed to be
Gaussianly distributed and can then enter a $\chi^2$ test. In the
following they will be denoted by $\sigma_{cb}$, $\sigma_{ub}$,
$\sigma_{\Delta m}$, $\sigma_\epsilon$, $\sigma_{\cal A}$
and $\sigma_R$ in an obvious notation. (The $\sigma_{\cal A}$ error is
related to the $\Delta m_{B_s}$ bound and is discussed separately below.)
For the quantities with Gaussian errors, we use \cite{Schn,PDG}
\beqa  \label{Eerrors}
|V_{cb}|&=&0.039\pm0.004,\nonumber \\
|V_{ub}/V_{cb}|_{\rm exp}&=&|V_{ub}/V_{cb}|_T\pm0.05,\nonumber \\
\Delta m_{B_d}&=&0.463\pm0.018\ {\mbox{ps}}^{-1},\\
|\epsK|&=&(2.258\pm0.018)\times10^{-3}.\nonumber
\eeqa
$|V_{ub}/V_{cb}|_T$ is a central value, defined below. (We actually use
yet another parameter in the fit, that is the top mass $\bar m_t$, with
the constraint $\bar m_t=165 \pm 8~ GeV$.)

A large part of the uncertainty in translating the experimental
observables to the CKM parameters comes, however, from errors related to
the use of hadronic models. In our work here these are related to the
value of $|V_{ub}/V_{cb}|_T$ (the subscript $T$ implies that we here
refer to the hadronic model dependent range for $|V_{ub}/V_{cb}|$ to
which an experimental error should be added to give the full uncertainty)
and to the parameters $B_{B_d}f_{B_d}^2$ and $B_K$ which enter the
calculations of $\Delta m_B$ and $\epsK$. At present, one cannot
assume any shape for the probability density of these quantities
(certainly not Gaussian) and include it in the fit. We thus do not
assume any shape for these distributions but use a whole set of
`reasonable' values for the parameters. Specifically, we scan the ranges
\beqa \label{Terrors}
0.06\leq|V_{ub}/V_{cb}|_T&\leq&0.10,\nonumber \\
160\leq f_{B_d}\sqrt{B_{B_d}}&\leq&240\ {\mbox{MeV}}, \\
0.6\leq B_K&\leq&1.0. \nonumber
\eeqa

Once a set of values $M=(|V_{ub}/V_{cb}|_T,f_{B_d}\sqrt{B_{B_d}},B_K)$
has been chosen, a classical least-square minimization can be performed
to estimate the CKM parameters (we use here the Wolfenstein
parameters \cite{WolPar} $A$, $\rho$ and $\eta$), by relating
measurements (with Gaussian errors) to theoretical calculations:
\beqa \label{chisquare}
\chi^2_M(A,\rho,\eta)&=&
\left({\langle|V_{cb}|\rangle-|V_{cb}|(A)\over\sigma_{cb}}\right)^2+
\left({|V_{ub}/V_{cb}|_T-|V_{ub}/V_{cb}|(\rho,\eta)
\over\sigma_{ub}}\right)^2 \nonumber \\
&+&\left({\langle\Delta m_{B_d}\rangle-\Delta m_{B_d}(A,\rho,\eta)
\over\sigma_{\Delta m}}\right)^2
+\left({\langle|\epsK|\rangle-|\epsK|(A,\rho,\eta)
\over\sigma_{\epsilon}}\right)^2,
\eeqa
where $\langle a\rangle$ denotes the experimental central value of
a quantity $a$.
To study the $(\rho,\eta)$ estimates obtained from the global fit,
we turn to the usual unitarity triangle representation. In this plane
we plot the hypercontour of $\chi^2=\chi^2_{\rm min}+5.99$ corresponding
to the 95\% CL contour. Sets of values $M$ with a $\chi^2$ probability
smaller than 0.05 are rejected and not shown in the plots.
Note that for each point of the contour
a new minimization is performed with respect to all parameters, meaning
that this method  takes into account the correlations between
the plotted parameters and all other ones. The superposition of the
contours for each scanned set of values $M$ is shown in our figures
together with the fitted estimates of $(\rho,\eta)$.

Also shown are the `minimum and maximum limit' contours obtained from
varying coherently all the uncertainties (theoretical uncertainties
are varied within the limits of (\ref{Terrors}) and experimental
errors between $2\sigma$). These last contours are just shown for
comparison, since their statistical meaning is not clear.

The $\chi^2$ can also be expressed in terms of another set of parameters:
$\chi^2(A,\sin 2\alpha,\sin 2 \beta)$. It is minimized in the same way
as before (using the 5\% probability cut) and the 95\% CL contours are
displayed in the ($\sin 2\alpha,\sin 2 \beta$) representation.
A subtlety that arises in this analysis is that of discrete ambiguities.
As a value $\sin2\phi$ ($\phi=\alpha$ or $\beta$) corresponds to
several possible values of $\phi$, there is a fourfold ambiguity
in the values of $(\rho,\eta)$ that correspond to a given pair
of values ($\sin2\alpha,\sin2\beta$). All four possibilities have
to be considered in the fit. In practice, two of them are always
incompatible with present data and consequently rejected by the
$P(\chi^2)>0.05$ cut.

\subsection{Including $\Delta m_{B_s}$ Properly}
The mass difference in the $B_s$ system has not been measured
and only 95\% CL limits have been obtained. Such a limit is only
a small part of the information and it cannot be included directly
in the $\chi^2$ minimization. These problems have been overcome
by the amplitude method that is now being used by the LEP
$\Delta m_{B_s}$ averaging Working Group \cite{Schn}.

For an initially ($t=0$) produced pure $B_s$, the probability
of a $\bar B_s$-tagging decay at time $t$ is
\beq
{\cal P}_m={1\over2\tau}e^{-t/\tau}(1-\cos\Delta m_{B_s}t),
\eeq
while that of a $B_s$-tagging decay at time $t$ is
\beq
{\cal P}_u={1\over2\tau}e^{-t/\tau}(1+\cos\Delta m_{B_s}t).
\eeq
($\tau$ is the $B_s$ lifetime.) The amplitude method
assumes that the probabilities are described by
\beq
{\cal P}_{m,u}={1\over2\tau}e^{-t/\tau}(1\pm{\cal A}\cos\Delta m_{B_s}t).
\eeq
Then, for each value of $\Delta m_{B_s}$, ${\cal A}$ and its uncertainty
$\sigma_{\cal A}$ are obtained. If ${\cal A}$ is compatible with 0,
there is no visible oscillation at this frequency. If ${\cal A}$ were
compatible with 1, an oscillation would be observed at this frequency.
The 95\% CL on $\Delta m_{B_s}$ is set at the frequency for which
${\cal A}+1.645\sigma_{\cal A}=1$.

To include this information in our fit, we calculate $\Delta m_{B_s}$ for
each set of the free parameters $(A,\rho,\eta)$ and find the
corresponding measured values of ${\cal A}$ and $\sigma_{\cal A}$.
This amplitude is then compared to the one expected if the tested
value of $\Delta m_{B_s}$ was the correct one (${\cal A}=1)$ and
the global $\chi^2$ is modified by adding
\beq
\left({{\cal A}-1\over\sigma_{\cal A}}\right)^2
\eeq
to the right hand side of eq. (\ref{chisquare}).

\subsection{Adding the FM Bound}

The FM bound is different from the other constraints that we use, in that
experiments give a measurement (see (\ref{CleoRange})) but the
clean information is only an upper bound (see (\ref{TheLimit})).
The way we implement this in our fit (based on a Maximum Likelihood
analysis) is the following. Suppose that the experimental result is
$R=\langle R \rangle\pm\sigma_R$. Then we add to the to
$\chi^2(A,\rho,\eta)$ a term of the form :
\beqa
0\ \ \ &{\rm if}&\ \ \ \sin^2\gamma(\rho,\eta)<\langle R \rangle,
\nonumber\\
\left({\sin^2\gamma(\rho,\eta)-\langle R \rangle\over\sigma_R}\right)^2\
\ \ &{\rm if}&\ \ \ \sin^2\gamma(\rho,\eta)>\langle R \rangle.
\eeqa

We also draw in the figures the two lines corresponding to 95\% CL
exclusion region which, for one-sided error, are given by
$\sin^2\gamma=R+1.645\sigma_R$.

\subsection{Including New Physics}

In the model independent analysis, New Physics effects can be
parameterized by 2 new parameters: $r_d,\theta_d$. The theoretical
calculations of $\Delta B=2$ processes are to be modified accordingly
\cite{GNW}:

\beqa \label{th_NP}
\Delta m_{B_d}^{NP} (A,\rho,\eta,r_d)=r_d^2 \Delta m_{B_d}^{SM}
(A,\rho,\eta)\nonumber \\
a_{\psi K_S}(A,\rho,\eta,\theta_d) = \sin 2(\beta(\rho,\eta)+\theta_d)
\\
a_{\pi\pi}(A,\rho,\eta,\theta_d) = \sin 2(\alpha(\rho,\eta)-\theta_d)
\nonumber
\eeqa

The full $\chi^2$ can then be written in terms of all the
unknown parameters, once enough measurements are available:
\beqa \label{chisquareNP}
\chi^2_M(A,\rho,\eta,r_d,\theta_d)&=&
\left({\langle|V_{cb}|\rangle-|V_{cb}|(A)\over\sigma_{cb}}\right)^2+
\left({|V_{ub}/V_{cb}|_T-|V_{ub}/V_{cb}|(\rho,\eta)
\over\sigma_{ub}}\right)^2 \nonumber \\
&+&\left({\langle\Delta m_{B_d}\rangle-\Delta m_{B_d}^{NP}
(A,\rho,\eta,r_d)\over\sigma_{\Delta m}}\right)^2 \\
&+&\left({\langle a_{\psi K_S}\rangle-a_{\psi K_S}(A,\rho,\eta,\theta_d)
\over\sigma_{a_{\psi K_S}}}\right)^2
+\left({\langle a_{\pi\pi}\rangle-a_{\pi\pi}(A,\rho,\eta,\theta_d)
\over\sigma_{a_{\pi\pi}}}\right)^2. \nonumber
\eeqa
In this case there is no extra degree of freedom to perform a $\chi^2$
probability test, but the minimization can be performed and contours
can be obtained.


{\tighten

}

\vfill\eject
\begin{figure}
\centerline{
\psfig{file=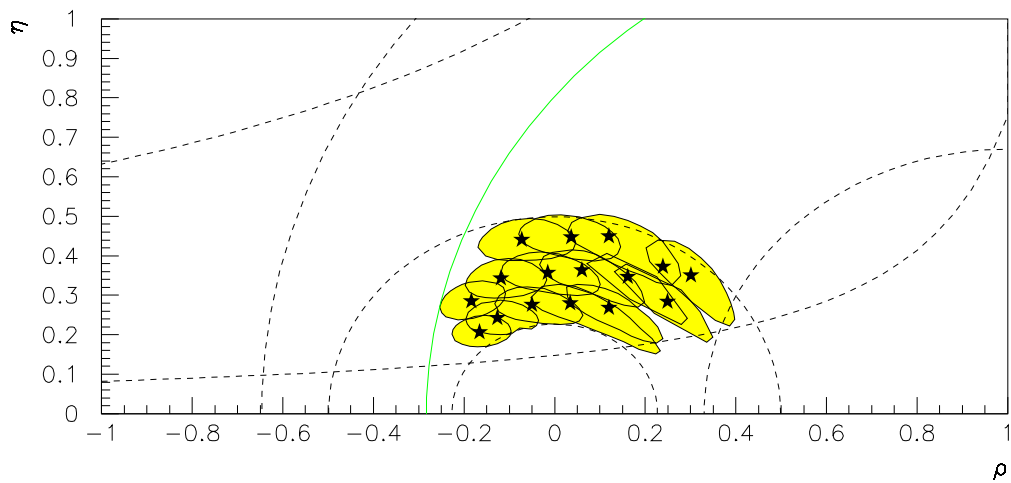,width=470pt,bbllx=0pt,bblly=350pt,bburx=612pt,bbury=792pt
 }}
\caption[a]{
The present allowed region for the Unitarity Triangle in the
$\rho-\eta$ plane. The input values  and the method used
for this determination are given in the Appendix.}
\end{figure}
\begin{figure}
\centerline{
\psfig{file=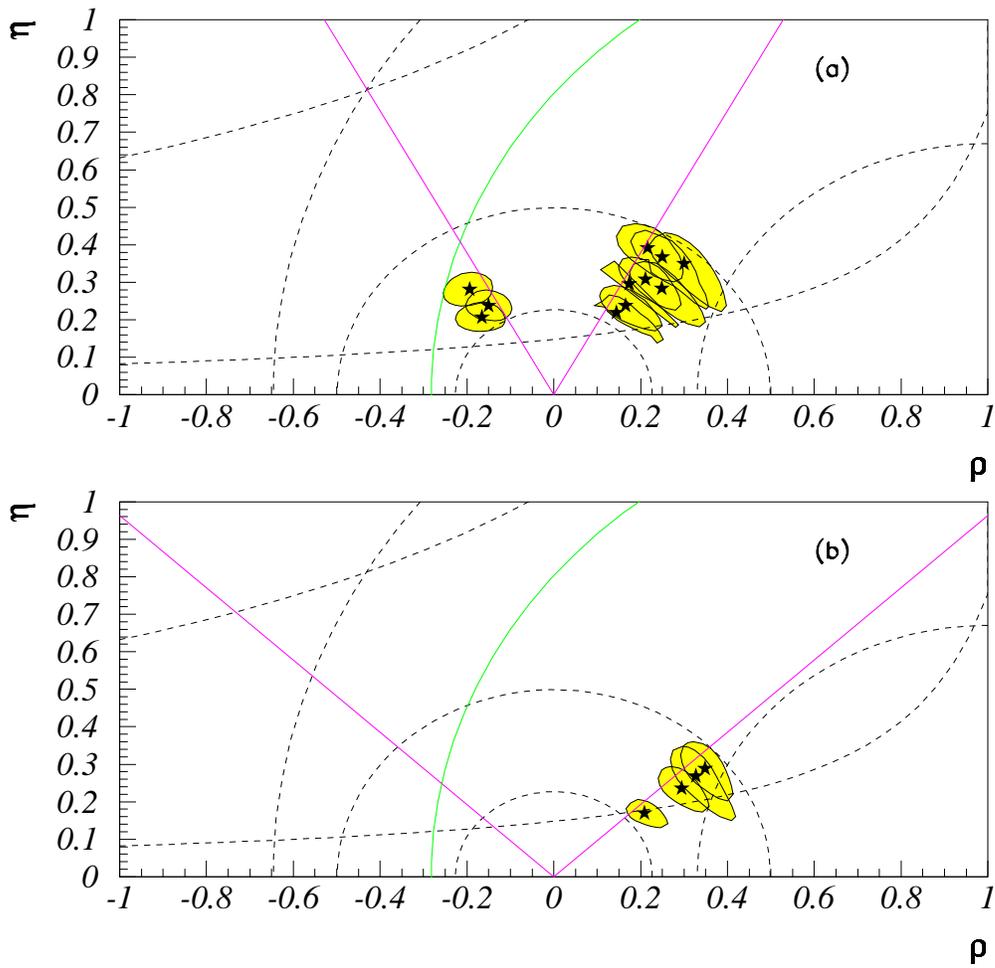,width=470pt,bbllx=0pt,bblly=90pt,bburx=612pt,bbury=792pt}
 }
\caption[b]{
The allowed range for the Unitarity Triangle for
$(a)$ $R=0.65\pm0.08$, $(b)$ $R=0.35\pm0.08$. For all other constraints
we use present data.}
\end{figure}
\begin{figure}
\centerline{
\psfig{file=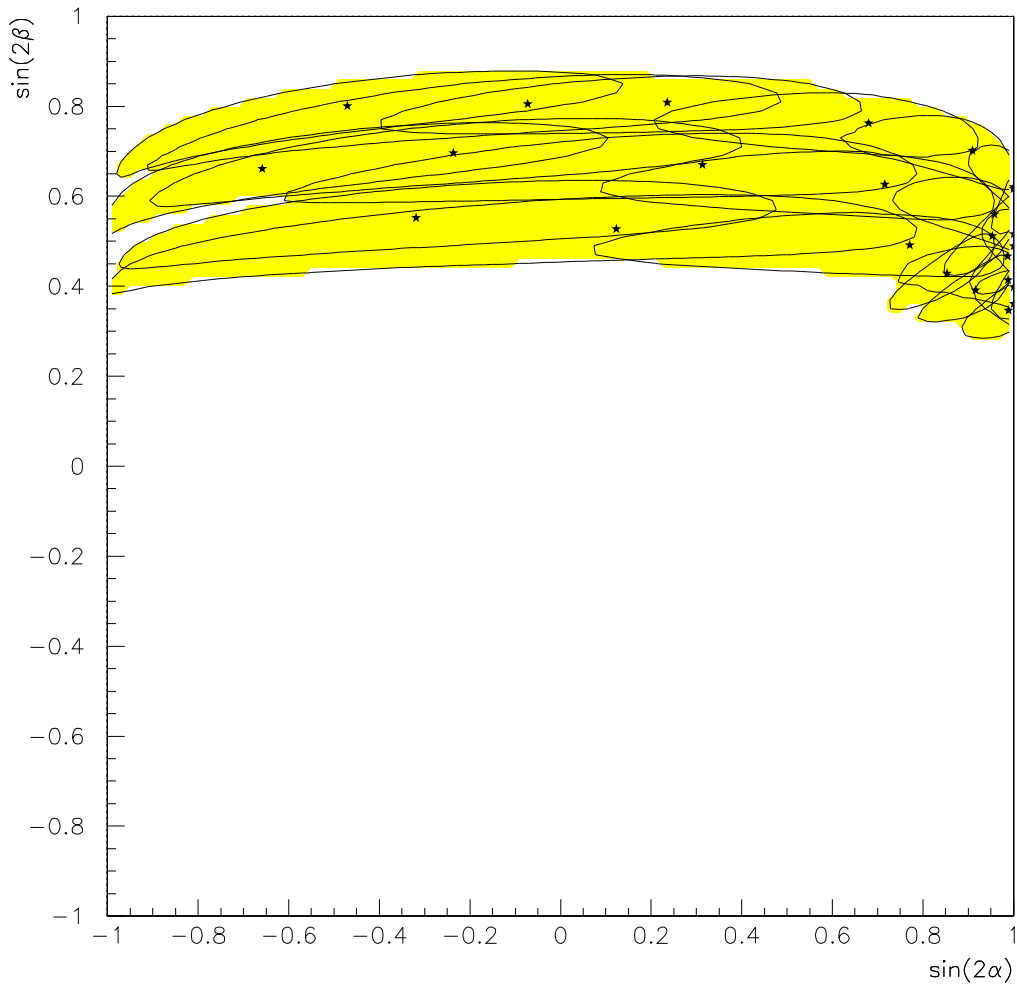,width=470pt,bbllx=0pt,bblly=90pt,bburx=612pt,bbury=792pt}
 }
\caption[c]{
The present allowed region in the $\sin2\alpha-\sin2\beta$ plane.
The input values and the method used for this determination are
given in the Appendix.}
\end{figure}
\begin{figure}
\centerline{
\psfig{file=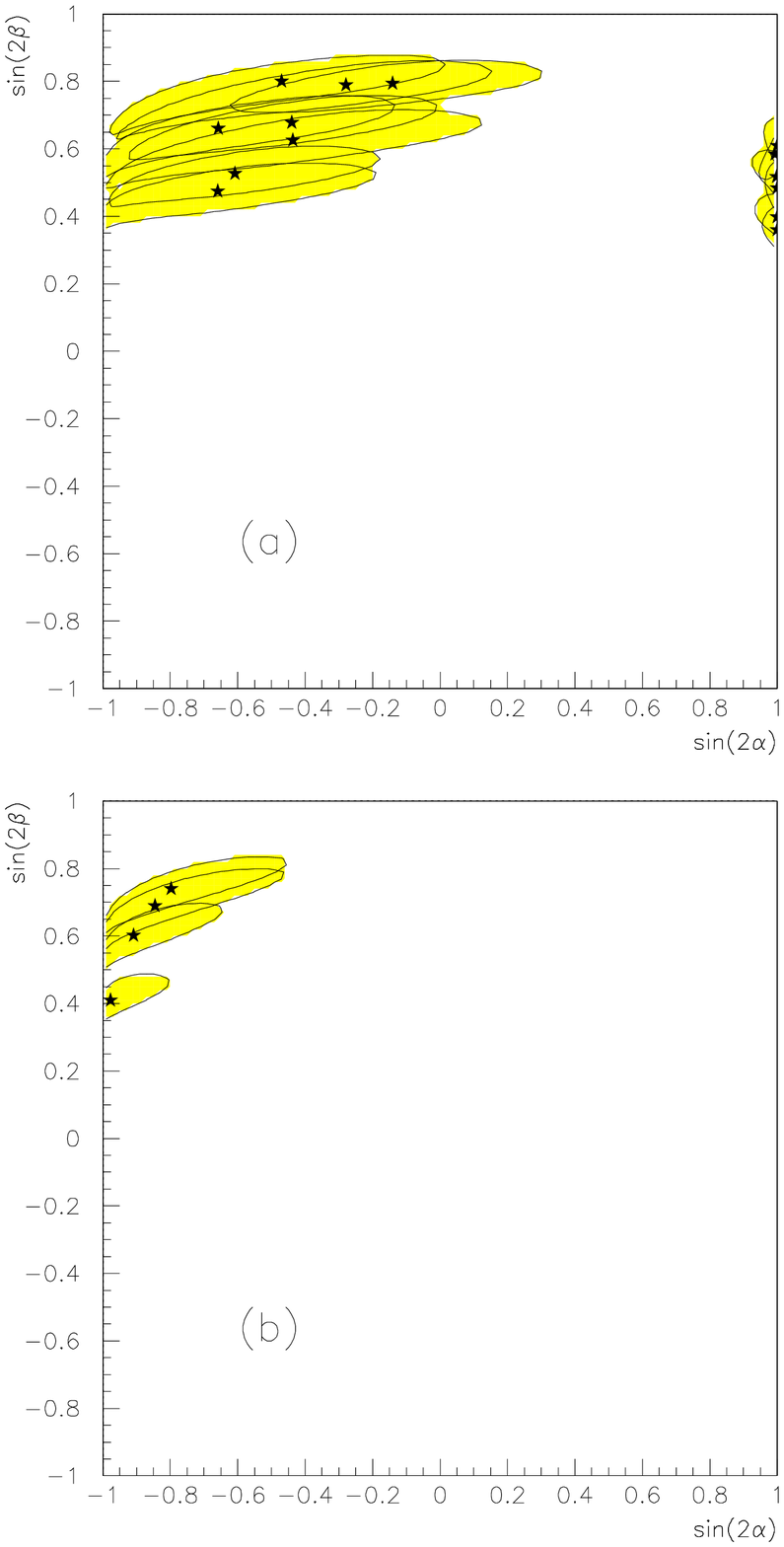,width=470pt,bbllx=0pt,bblly=10pt,bburx=612pt,bbury=792pt}
 }
\caption[d]{
The allowed range in $\sin2\alpha-\sin2\beta$ plane for
$(a)$ $R=0.65\pm0.08$, $(b)$ $R=0.35\pm0.08$. For all other constraints
we use present data.}
\end{figure}
\begin{figure}
\centerline{
\psfig{file=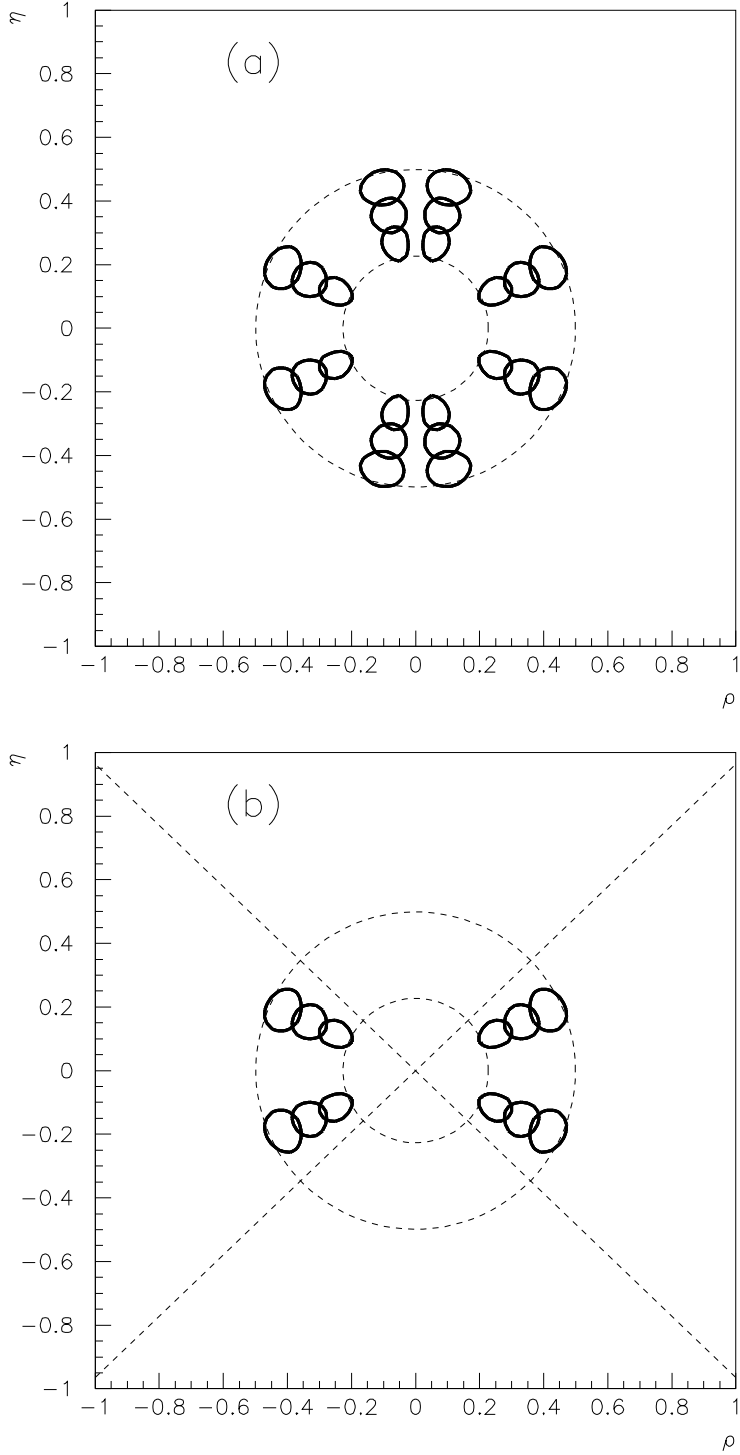,width=470pt,bbllx=0pt,bblly=10pt,bburx=612pt,bbury=792pt}
 }
\caption[e]{
The model independent construction of the Unitarity Triangle
$(a)$ without the FM bound and $(b)$ with $R=0.35\pm0.08$.
We use the current range for $|V_{ub}/V_{cb}|$,
$\aPP=0.60\pm0.09$ and $\aPK=0.20\pm 0.06$.}
\end{figure}
\begin{figure}
\centerline{
\psfig{file=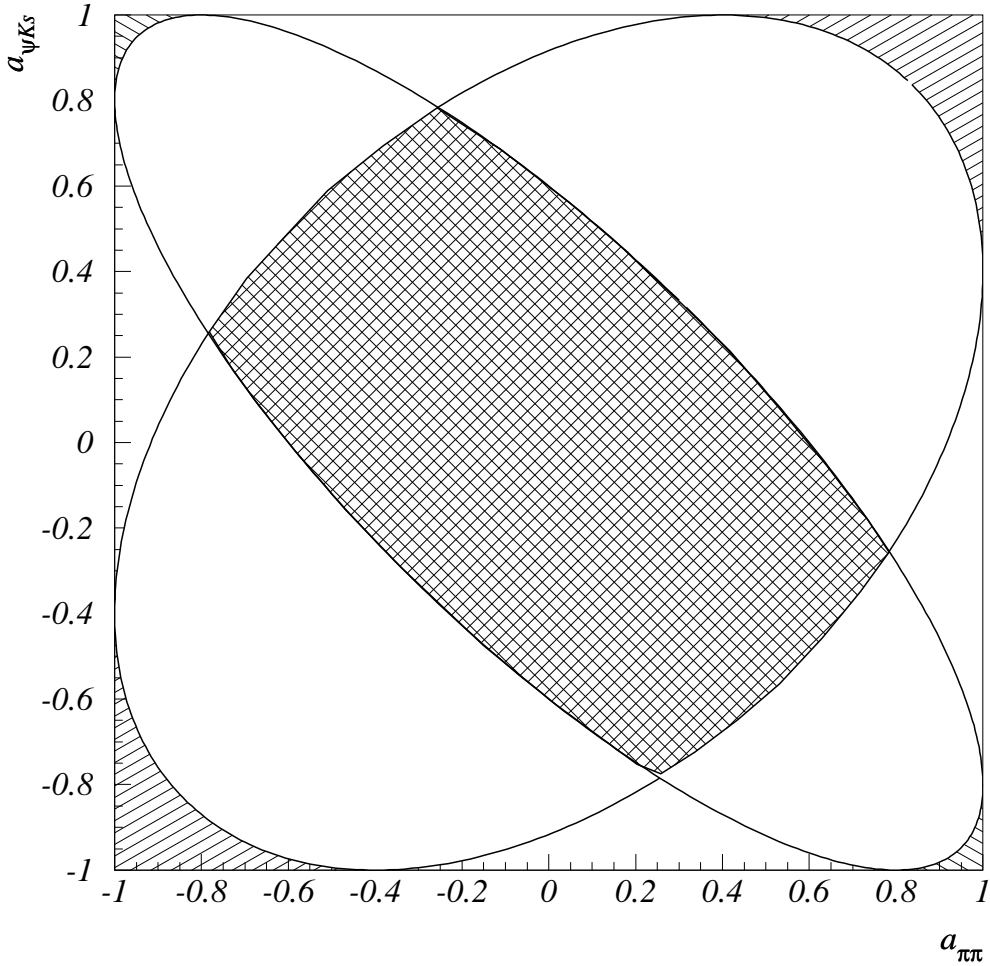,width=470pt,bbllx=0pt,bblly=70pt,bburx=612pt,bbury=792pt}
 }
\caption[f]{
The allowed range in the $\aPP-\aPK$ plane with New Physics
satisfying the conditions specified in the text. The right-hatched
area is excluded by an upper bound $\sin^2\gamma<0.7$. The left-hatched
area is excluded by a lower bound $\sin^2\gamma>0.1$.
For the combined bound, $0.1<\sin^2\gamma<0.7$, the cross-hatched
area is also excluded.}
\end{figure}
\end{document}